# The Impact of Multitasking in Video Meetings on Meeting Outcomes: Initial Results

*Short Paper*

Julia Seitz, Karlsruhe Institute of Technology, Germany, julia.seitz@kit.edu

Fabio Stano, Karlsruhe Institute of Technology, Germany, fabio.stano@kit.edu

René Riedl, University of Applied Sciences Upper Austria, Austria; Johannes Kepler University Linz, Austria, rene.riedl@fh-steyr.at

Alexander Maedche, Karlsruhe Institute of Technology, Germany, alexander.maedche@kit.edu

## Abstract

Multitasking in video meetings is a ubiquitous phenomenon. While extant literature has investigated the circumstances under which people multitask in video meetings, little is known about the impacts of multitasking on meeting outcomes. Building on the findings of previous studies investigating the impact of multitasking in individual non-video meeting work setups and threaded cognition theory, we present a research model that suggests that multitasking may have negative effects on meeting outcomes such as process loss and team meeting outcome satisfaction. We tested the research model in an incentivized within-subject laboratory experiment with n=72 participants. Results suggest that multitasking has a negative effect on process loss and team meeting outcome satisfaction compared to non-multitasking. We plan to further investigate the mediating role of cognitive user states and neurophysiological activations and ultimately aim to contribute with a deeper understanding of the impacts of multitasking.

## 1 Introduction

A key challenge of video meetings is a greater number of potential distractions that users are faced with, potentially leading them to appear to be more absent (Showpad, 2022). [1] Frequently, these distractions stem from engaging in outside activities, such as checking emails, processing other individual tasks, or browsing the internet (Cao et al., 2021). While some activities might be related to the meeting (e.g., writing meeting minutes), Cao et al. (2021) show that approximately 30% of video meetings involve activities such as writing (unrelated) mails in parallel to meetings, resulting in multitasking.

[1] Following the AIS Professional Conduct, the authors disclose that parts of the paper content are part of an unpublished PhD thesis

Multitasking is defined as doing multiple activities at the same time (Koch et al., 2018). Current literature on multitasking in video meetings reports the extent of multitasking, performed activities, and their impact on social factors or learning performance (e.g., Cameron & Webster (2013), Chang et al. (2022)). However, only very few studies investigate the actual effects of in-meeting multitasking on the participants' cognitive resources and meeting outcomes in the work context. This is in line with Cao et al. (2021, p. 2)'s claim that "[extant works] are mainly based on small-scale qualitative studies [...]. As a result, there is little statistical evidence […] how such remote multitasking behavior may affect workers and groups". This lack of research is a serious problem, given how common multitasking is and the negative impact in individual work contexts (Cao et al., 2021; Koch et al., 2018). Specifically, we argue that the transferability of the findings of the individual work context to the meeting context is not clear. First, a key difference between a video meeting and the individual work setup lies in the social dimension, which may influence multitasking behavior and effects. Driven by social desirability, multitasking may be increasingly effortful, when aiming to be performed secretly, thinking about whether it may be obvious to others (Suh et al., 2018). On the other hand, the virtuality of the video meeting allows one to hide behind the screen, reducing aspects of social impacts (Suh et al., 2018). Hence, it is unclear how virtuality and social dynamics interact, raising the question of whether findings from individual multitasking hold true in meetings. Second, outcomes in meetings are more complex than in individual work. Unlike individual tasks, the outcome is influenced by the overall team, not only the single individuum. Hence, meeting outcomes comprise several aspects, amongst meeting performance and satisfaction (Briggs et al., 2006; Reinig, 2003). Both can exist when people commit to the outcome and feel like it is of a high quality, even though it is not based on their own opinion and efforts (Green & Taber, 1980).

To conclude, we emphasize the relevance of multitasking as a frequent behavior and argue that from a theoretical standpoint, the impact of multitasking in video meetings on outcomes remains underspecified and it is unclear whether and how findings from individual work contexts transfer to the team meeting setup (Cao et al., 2021; Janssen et al., 2015). To close the research gap, we report a laboratory study to answer the question of what is the impact of multitasking in video meetings on meeting outcomes?

Accordingly, we build and test a research model. Grounded in theory of threaded cognition (Salvucci & Taatgen, 2008), we suggest that distraction leads to additional cognitive load and ultimately to negative work outcomes. Following Good et al. (2016), Reinig (2003), and Sarker et al. (2019), we distinguish the meeting outcomes into instrumental outcomes, conceptualized as process loss, and outcomes contributing to instrumental and social dimensions, conceptualized as team meeting outcome satisfaction. We present initial results and the model below.

## 2 Conceptual Foundations

Multitasking refers to performing multiple activities at once, involving multiple information-processing activities happening in overlapping time frames (Koch et al., 2018). While not researching multitasking explicitly, existing IS literature investigated the effects of elements that trigger multitasking, such as interruptions (Nadj et al., 2023; Stangl & Riedl, 2023). Moreover, multitasking is often triggered by self-interruptions (e.g., when being bored) or mixed interruption cues (Adler & Benbunan-Fich, 2013; Jin & Dabbish, 2009). While it is discussed whether multitasking is a form of an interruption, unrelated, or a result of it, we follow Cao et al. (2021) and argue that it is an outcome (Puranik et al., 2020). Studies on multitasking in meetings investigate the rationale for multitasking and type of activities while multitasking (e.g., writing mails and individual work as the most mentioned types of activities (Cao et al., 2021), multicommunication (Cameron & Webster, 2013; Dennis et al., 2010)). Besides, studies focus on the impact of organizational factors (e.g., hierarchy, timing of call (Cao et al., 2021), norms (Stephens & Davis, 2009)), task type (De Bruin & Barber, 2022), or VMS design factors (Sarkar et al., 2021), and how they shape multitasking behavior and its impact. For outcomes, Twyman et al. (2020) investigate decreases in manipulation of meeting participants, and Chang et al. (2022) report effects on learning outcomes. Studies mostly collect data in surveys and ethnographic studies, limiting the extraction of the detailed impacts that specific stimuli have. Experimental studies are limited to learning and social outcomes (Chang et al., 2022; Tchanou et al., 2021). However, compared to the passive lectures, work meetings require active participation and have few participants, both driving alterations in multitasking (Cao et al., 2021).

Video meetings and their tools have been studied in IS for decades. The virtual layer mediating the communication provides a higher possibility of hiding in contrast to face-to-face meetings, since the individual with whom one is conversing is unable to discern what one is viewing on the screen (Dennis et al., 2010; Suh et al., 2018). Consequently, it is more challenging for other meeting participants to identify one's current activity, such as the opened application window on the screen (De Bruin & Barber, 2022). Meetings are often held to connect workers resulting in conversations, to exchange information, or solve a problem (Abramova et al., 2024). Consequently, these meetings lead to different outcomes. Similar to outcomes at a workplace in general, they can be differentiated into instrumental work-related, and more social-or relationship-related outcomes (Good et al., 2016; Reinig, 2003; Sarker et al., 2019). For meetings, instrumental outcomes center around objective and subjective performance in the meeting and encompass meeting productivity. While most studies refrain from measuring the pure task performance as the meeting outcome variable (Abramova et al., 2024), an alternative approach stemming from multitasking investigates the degree to which individuals believe to have missed relevant parts of the meeting, referred to as process loss (Cameron & Webster, 2013). Social dimensions target satisfaction with the team and meeting (Green & Taber, 1980; Reinig, 2003).

## 3 Research Model

To capture the complexity of the multidimensional outcomes, our study focuses on two key dimensions: process loss as a work-related instrumental outcome measure, and team meeting outcome satisfaction -as part of meeting satisfaction -as an outcome contributing to instrumental and social dimensions. In multitasking research, several measures for performance outcomes exist (Benbunan-Fich et al., 2011). Combining work on meeting outcome measures and multitasking, we measure the loss of the primary task when engaging in the secondary task as a metric to evaluate the individual's meeting performance, an instrumental work-related outcome. Process loss can be defined as "the series of actions, behaviors, or states experienced during the multitasking episode that inhibit the focal individual from accomplishing his or her communication tasks" (Cameron & Webster, 2013, p. 357). In multitasking scenarios involving conversations, process loss often manifests as missing parts of the discussion, necessitating repetition. A common example is when an individual is suddenly addressed in a meeting but has lost track of the conversation. Given the premise that individuals are unable to maintain focus and commitment to a single task when they are engaged in multitasking, we expect that multitasking results in a higher process loss than non-multitasking and propose: H1a: Multitasking in video meetings leads to higher process loss compared to non-multitasking.

As a second dimension, we consider satisfaction with the team meeting outcome as an essential outcome that has shown to significantly impact success in practice (Niederman et al., 1996). While there is a debate about team and meeting satisfaction, we follow Reinig (2003) and define team meeting satisfaction to consist of two elements: process satisfaction and outcome satisfaction. Green & Taber (1980) assess the individual's satisfaction with the team outcome and how much the individual commits to the solution. Given that compared to an individual setup the individual can commit to the solution without answering -when trusting the rest of the team, team meeting outcome satisfaction may be present even though the individual contribution is small. Reinig (2003) propose that when not contributing much, satisfaction may be small when the goal was to actively participate in the meeting due to an under fulfillment of the goal to actively participate, depending on the set goal. However, if the goal is simply to attend the meeting while focusing on individual tasks, this goal may be fulfilled, leading to satisfaction with the meeting and its outcome, even in the presence of multitasking. This may explain reported productivity increases in multitasking (Cao et al., 2021). We note that perceived goal achievement is a key factor in determining satisfaction, and satisfaction tends to increase when the achieved goal is more challenging (Briggs et al., 2006; Reinig, 2003). Hence, achieving satisfaction due to reaching the lower goal to "sit in the meeting" can exist, but should be less satisfying than the goal of participating. Transferred to multitasking, we hypothesize: H1b: Multitasking in video meetings leads to lower team meeting outcome satisfaction compared to non-multitasking.

# 4 Experimental Design

We employ a counterbalanced cross-over within-subject study design including two conditions (multitasking, non-multitasking), to mitigate carry-over effects. We opted for a within design after reflecting validity, causality, power and participant effort due to sensor data collection (Montoya, 2023).

**Manipulation and Tasks:** To simulate realistic multitasking behavior with both tasks visible at once, we created a meeting page with a split screen (Figure 1). On the left part of the screen, the video meeting including the team members' images and the team task was displayed. On the right part of the screen, the individual task was shown. Participants in the non-multitasking condition saw a blank space instead of the individual task. Participants were informed about being an office worker in a fictional company, that needs to participate in top talent group, including regular team building meetings (i.e. meeting in our setup). In addition, when being assigned to the multitasking condition, they were informed about a system breakdown that required additional manual office work, presented in the form of the secondary task, an individual invoice matching task. In the meeting task, participants worked on a group decision-making task (Mc Grath, 1984), the NASA survival on the moon task and its equivalent of the lost at sea task (Hall & Watson, 1970; Knox, 2022). Both tasks are used for team building activities (Seitz et al., 2024) and a sample solution exists. Based on the introductory scenario, participants were required to solve a task that covered the matching of several invoices to a payment as individual task (only in multitasking). This task resembles traditional office work in organizations (Nadj et al., 2023).

**Incentives:** We used balanced incentives of the meeting and the individual task with both tasks leading to the same possible payout, to avoid preferences for a task. Total payout of the experiment consisted of fixed payment (5€), 4€ for solving the questionnaires, and the meeting payout (up to 8€ per meeting, based on % correctness of the team solution with the sample solution, and up to 8€, depending on the number of correct invoice matching tasks for the individual task). We adjusted the payouts in pre-studies to ensure a compensation in line with the laboratory payout rules (minimum wage compensation).

**Participants:** We recruited via a large student panel from a public university. Within one experimental session, participants were randomly assigned to the multitasking or non-multitasking condition. Each team consisted of five participants. One participant was assigned as the team leader, being in charge of creating the team's ranking of items on the shared screen. To avoid influences of team role, this team leader was excluded in the analysis. We recruited 90 participants with 72 participants being exposed to the experimental conditions (sample size: n = 72). Gender was balanced, participants were on average 24.74 years old (SD=4.33), and not familiar with others prior to the study. Participants reported familiarity with video meetings (M=5.29, SD=1.68), equal experience with the topics of task (sailing: M=1.94, SD= 1.58, astronautics: M=1.90, SD=1.25), and a medium-level experience with arithmetic tasks (M=3.97, SD=1.76) on a 7-point Likert scale.

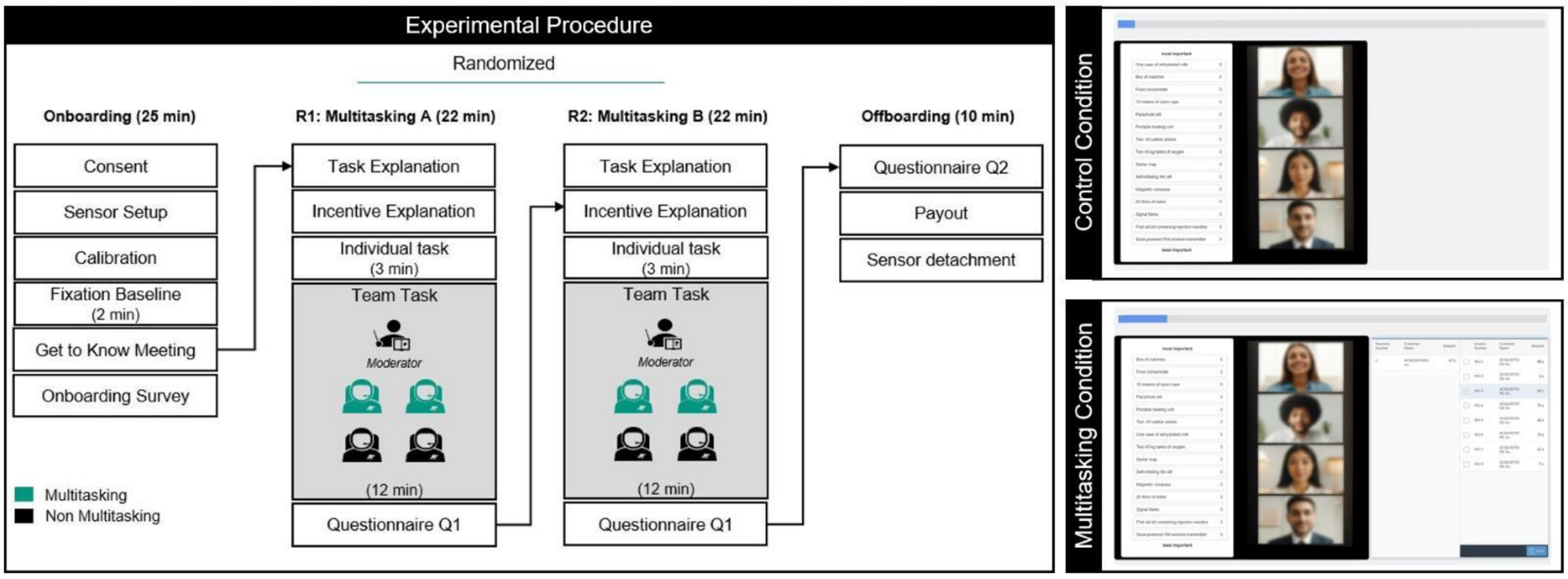

*Figure 1. Experimental Procedure and Treatments.*

**Procedure:** We received data protection and ethics approval and pre-registered the study (asPredicted #183660). During the onboarding phase, sensors to record neurophysiological data were attached and calibrated. Participants then read the instructions. Afterward, participants filled out the onboarding survey (demographic questions). To minimize familiarity bias, participants took part in a five-minute get-to-know meeting. Then, participants began the working phase, comprised of two rounds, with instructions, the individual and team working phase, and a meeting questionnaire each. In each round, participants were informed about the current scenario of their treatment (i.e., working in a company, participating in a team building meeting, required to work on an additional task or not), as well as an explanation of the incentivization. Participants in the multitasking treatment received an explanation of the secondary invoice task and practiced the task for three rounds. Afterward, participants were directed to a meeting with all participants (12 minutes) and discussed their individual opinions on the task to find a joint solution. Depending on the treatment, participants were given the opportunity to work on the secondary invoice matching task or not (Figure 1). After each meeting, participants were navigated to a survey about our constructs of interest (process loss, team meeting outcome satisfaction) and user states (cognitive load, flow). This procedure was repeated for the second round and the treatment conditions were switched. In the offboarding phase, participants completed the final survey (control factors).

**Data collection:** We measured outcomes as process loss (adapted from (Cameron & Webster, 2013)), and meeting outcome satisfaction (Green & Taber, 1980). We collected user states of cognitive load (Hart & Staveland, 1988) and flow (Engeser & Rheinberg, 2008), demographic data (i.e., gender, education, familiarity between participants, age (own scales)), and personality traits as control variables. We measured traits relevant to meeting behavior (Big5 (Rammstedt et al., 2014), communication apprehension (McCroskey, 2015), topic experience (own)) and related to multitasking (monochronic-polychronic tendency (Lindquist & Kaufman-Scarborough, 2007)). To confirm the manipulation, we measured the ability to work on meeting related activities (own, 7-item Likert). We collected continuous user data via heart rate (ECG, biosignalplux), brain activity (OpenBCI), and eye-tracking data (Tobii). This analysis targets meeting outcomes, we report further collected data for transparency reasons.

## 5 Preliminary Results

For our preliminary analysis, we checked the reliability and validity of the measurement scales via Cronbach alpha and item-total correlations in each subscale, ensuring sufficient construct loadings (higher than 0.4) and Cronbach alpha levels (higher than 0.7) (Nunnally & Bernstein, 1994). We averaged all items per construct and checked the normality of the data distribution using Shapiro-Wilk tests identifying non-normal distributed data. We checked for carry-over effects in our within-subject design but found none. For testing the distinct differences, we used paired Wilcoxon signed rank tests, corrected with the Holm Method for testing of multiple constructs (Benjamini & Yekutieli, 2001).

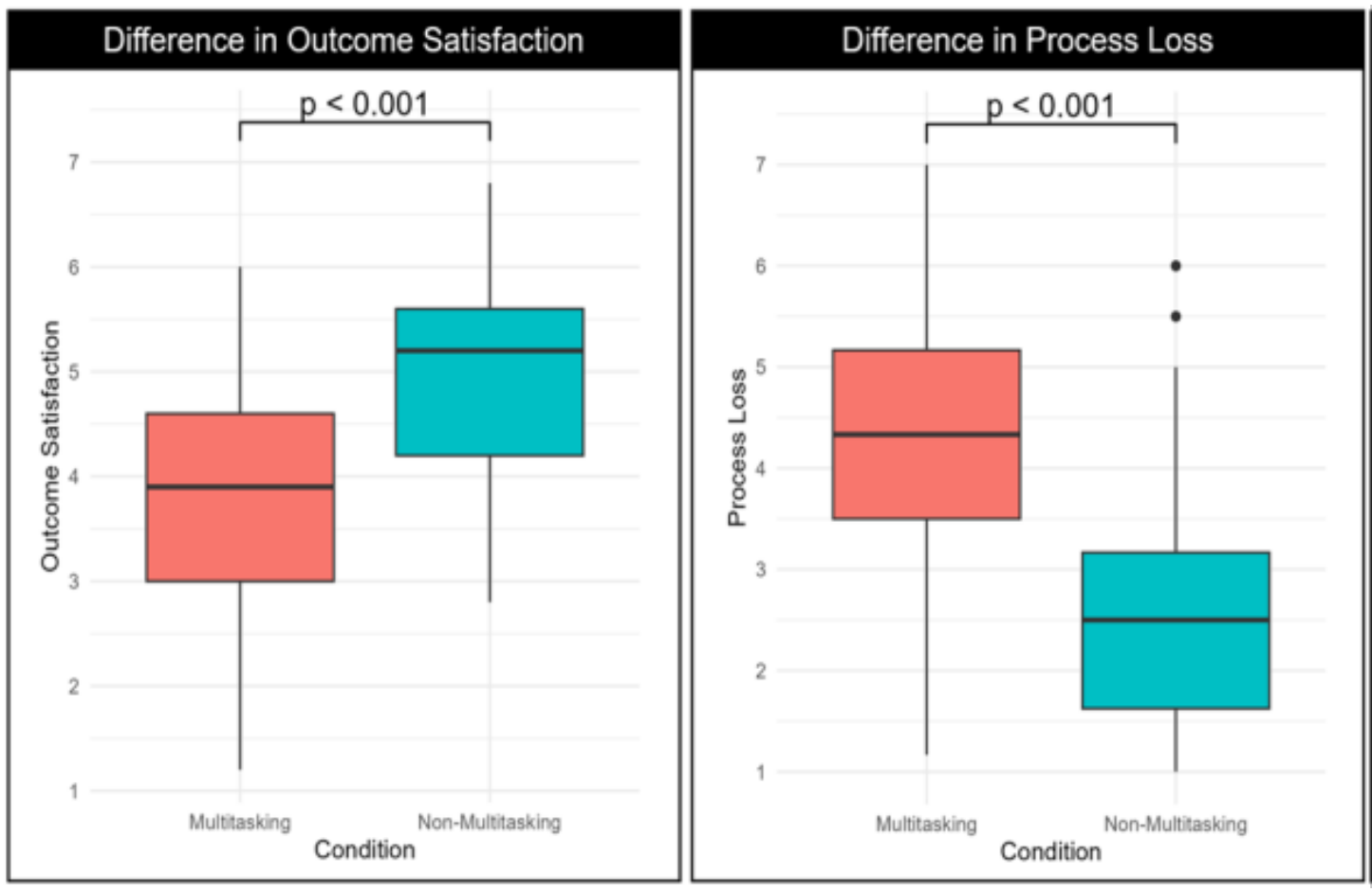

Descriptive Statistics

| | | Multitasking | Non-Multitasking |
|---|---|---|---|
| Outcome Sat. | Mean | 3.84 | 4.89 |
| | SD | 1.10 | 0.99 |
| | Median | 3.9 | 5.2 |
| | IQR | 1.6 | 1.4 |
| Process Loss | Mean | 4.22 | 2.56 |
| | SD | 1.38 | 1.16 |
| | Median | 4.33 | 2.5 |
| | IQR | 1.67 | 1.54 |

*Figure 2. Comparison of Meeting Outcome Satisfaction and Process Loss Between Treatments.*

**Manipulation Check:** Participants reported a higher agreement to working only on meeting relevant activities in the non-multitasking condition compared to the multitasking condition (M=4.58 SD=0.9, Median=5.0 for non-multitasking, M=1.67, SD=1.23, Median=1.0 for multitasking). The manipulation check confirmed the success of manipulation with significantly different values ($V = 8.5$, $p < .001$).

**Effects of Multitasking vs. No-Multitasking:** We identified support for hypothesis H1a, as the results show significantly higher process loss when multitasking ($V = 286$, $p < 0.001$, confidence interval 95% CI = [-2.25, -1.33], effect size e = 0.617). Taking a look at team meeting outcome satisfaction, we found significantly lower team meeting outcome satisfaction when multitasking compared to non-multitasking, supporting hypothesis H1b ($V = 2135.5$, $p < 0.001$, CI = [0.80, 1.40], e = 0.617).

## 6 Discussion and Concluding Remarks

Our preliminary findings suggest that multitasking results in higher process loss, and lower team meeting outcome satisfaction, supporting H1a und H1b. Negative effects on the process loss in the meeting are comparable to Cameron & Webster (2013). While Cameron & Webster (2013) argue that process loss typically increases with more than two (communication) tasks, our findings show that two non-communication tasks already lead to process loss. Findings further suggest that satisfaction is significantly decreased when multitasking, showing effects for mixed instrumental-social-oriented outcomes, previously less investigated. As two out of five team members were multitasking in our study, this raises the question of whether lower satisfaction exists when multitasking oneself compared to when other team members are multitasking. We see an explanation for the large effect size in the core difference of our study from previous individual-work: the social setting. Feeling observed and related effort on understanding how one needs to present oneself might be in the back of the mind of the individual and lead to more secretly engaging in multitasking, worsening its impact. Similar effects have been shown for fatigue (Bailenson, 2021). This serves as a starting point for the next phase of our research.

To better understand mechanisms that can explain the impact of multitasking, we analyze cognitive user states and their mediating roles according to the stimulus-organism-response model (Mehrabian & Russell, 1974) to further specify the nomological net of multitasking in the context of video meetings and better understand the actual reasons for the increased process loss and reduced satisfaction. In addition, we aim to investigate the neurophysiological correlates of the user states to understand bodily reactions and gaining a better understanding of virtual collaboration through a NeuroIS lens. In this regard, investigating boundary conditions, such as task complexity or individual differences in polychronicity, could be of interest for future research.

A second explanation could stem from the incentivization scheme. By adjusting the incentive structure, future research may better understand how other differing demands influence multitasking (high incentivization of the individual task to symbolize deadlines, or high incentives on meeting to show high-stakes meetings). Additionally, investigating the impact of multitasking (e.g., the number of multitaskers in a team) on team dynamics (e.g., active participation, cohesion) is a worthwhile direction for future studies. We contribute to the call from Janssen et al. (2015) and IS research on e-collaboration and users' response to technology (Agogo & Hess, 2018), as a we explore the impact of technology mediation on communication, and extend work on technology-driven interruptions (Stangl & Riedl, 2023) by investigating multitasking as a result of technology-and user-driven interruptions. We contribute to practice as our findings can inform remote workers and organizations when creating trainings and meeting guidelines that help to promote more sustainable workplace behaviors that reduce negative work outcomes and may serve as a call to rethink organizational factors that facilitate and encourage multitasking. After analyzing the user states, we hope to support designers in creating assistance systems that show awareness interventions.

These contributions are subject to the limited ecological validity of laboratory studies (Balapour & Riedl, 2025). We suggest conducting studies with different setups, including other team constellations (e.g., ongoing working teams), tasks (esp. comparison of writing vs. analytical task), or incentives. To conclude, we address the research gap of understanding how multitasking impacts meeting outcomes, revealing that multitasking significantly deteriorates process loss and satisfaction. Next, we analyze mechanisms that explain the negative effects on outcomes, by exploring the role of cognitive user states and neurophysiological correlates. We encourage participants to reflect on their multitasking habits and urge designers and organizations to explore assistance systems, training, and guidelines to enable responsible multitasking.

**Acknowledgments.** Funded by the Deutsche Forschungsgemeinschaft (DFG, German Research Foundation)-GRK27391-Project Nr. 447089431-Research Training Group: KD School-Designing Adaptive Systems for Economic Decisions.